\documentclass[12pt]{article}

\def\be{\begin{equation}}
\def\ee{\end{equation}}

\newtheorem{theorem}{Theorem}
\newtheorem{lemma}[theorem]{Lemma}

\newtheorem{definition}[theorem]{Definition}
\newtheorem{proposition}[theorem]{Proposition}

\begin{document}

\title{Ultrametric pseudodifferential operators \\ and wavelets for the case \\ of non homogeneous measure}

\author{S.V.Kozyrev\footnote{Steklov Mathematical Institute, Moscow, Russia}}

\maketitle

\begin{abstract}
A family of orthonormal bases of ultrametric wavelets in the space
of quadratically integrable with respect to arbitrary measure
functions on general (up to some topological restrictions)
ultrametric space is introduced.

Pseudodifferential operators (PDO) on the ultrametric space are
investigated. We prove that these operators are diagonal in the
introduced bases of ultrametric wavelets and compute the
corresponding eigenvalues.

Duality between ultrametric spaces and directed trees is
discussed. In particular, a new way of construction of ultrametric
spaces by completion of directed trees is proposed.
\end{abstract}

\section{Introduction}

In the present paper we continue to develop the analysis of
pseudodifferential operators on ultrametric spaces, following
\cite{IZV}.

We consider for an ultrametric space $X$ the directed tree ${\cal
T}(X)$ of balls in $X$, and consider the partial order on
$X\bigcup {\cal T}(X)$ defined by inclusion of balls and inclusion
of points of $X$ into balls. We consider on the space $X$ a
($\sigma$--additive and possessing a countable or finite basis)
measure $\nu$ of a general form and investigate pseudodifferential
operators of the form \be\label{opT} Tf(x)=\int T^{({\rm
sup}(x,y))}(f(x)-f(y)) d\nu(y) \ee acting in the space
$L^2(X,\nu)$ of quadratically integrable complex valued functions.
The integration kernel $T^{(I)}$ is a function on the tree ${\cal
T}(X)$ and $I={\rm sup}(x,y)$ is defined by the partial order on
$X\bigcup {\cal T}(X)$.

We introduce the orthonormal basis of ultrametric wavelets in the
space $L^2(X,\nu)$, which diagonalizes the pseudodifferential
operator $T$, and compute the corresponding eigenvalues.

Also we discuss the duality between ultrametric spaces and
directed trees. In particular, we propose a new construction of
ultrametric space as a completion of a directed tree ${\cal T}$
(i.e. of a tree with a direction, or defined in the special way
partial order) with respect the the metric, defined by the
direction.

The present paper develops the approach of the paper \cite{IZV},
where the particular case of the analysis ultrametric
pseudodifferential operators of the type (\ref{opT}) was
considered, for which the measure $\nu$ was chosen in the special
way, for which the maximal subballs in an ultrametric ball have
the equal measure. In the present paper we consider more general
case, for which the measures of the balls can be arbitrary
positive numbers. We call this case the case of non homogeneous
measure. Moreover, in \cite{IZV} the more standard construction of
ultrametric space as the set of classes of equivalence of
decreasing infinitely continued paths in the directed tree was
applied.

Investigation of ultrametric pseudodifferential operators was
started \cite{Vl1} with the introduction of the Vladimirov
operator of $p$--adic fractional derivation, see \cite{VVZ} for
detailed exposition. Different problems of $p$--adic analysis and
$p$--adic mathematical physics were considered in
\cite{Andr3}--\cite{Kochubei2}.

The Vladimirov operator can be diagonalized by the $p$--adic
Fourier transform. Also there exist bases of eigenvectors with
compact support \cite{VVZ}. In paper \cite{wavelets} the basis of
$p$--adic wavelets in the space $L^2(Q_p)$ of quadratically
integrable complex valued functions on the field of $p$--adic
numbers was introduced and it was shown that this basis is a basis
of eigenvectors of the Vladimirov operator. In papers
\cite{Benedetto}, \cite{Benedetto1} the construction of the
wavelet basis of \cite{wavelets} was generalized onto more general
local fields and groups.

In paper \cite{nhoper} the family of pseudodifferential operators
in the space $L^2(Q_p)$, diagonal in the basis of $p$--adic
wavelets, but not diagonalizable by the Fourier transform, was
constructed, and the corresponding eigenvalues were computed.
Further generalization of this result onto the case of
pseudodifferential operators in $L^2(Q_p)$ of more general form
was performed in \cite{trudy}, \cite{0403440}. In paper \cite{IZV}
theory of ultrametric wavelets and PDO related to general
ultrametric spaces was developed (in less general case, compared
to the present paper).

Theory of ultrametric pseudodifferential operators has physical
applications. In papers \cite{ABK}, \cite{PaSu} it was shown that
the Parisi matrix, which describes replica symmetry breaking is a
discrete analogue of some $p$--adic pseudodifferential operator.
In papers \cite{ABKO}, \cite{Trudy2} relation between ultrametric
diffusion and dynamics of macromolecules was discussed. For the
review of the results of $p$--adic mathematical physics see
\cite{VVZ}. In particular, one can mention applications to string
theory \cite{Vstring}, \cite{Freund}, and mathematical models of
biology and cognitive science \cite{Andr1}, \cite{Andr2}.

The present paper has the following structure.

In Section 2 build the structure of a directed tree on the set
${\cal T}(X)$ of balls in an ultrametric space $X$.

In Section 3 we introduce the family of directed trees and define
ultrametric on the trees from this family.

In Section 4 we perform completion of the trees with respect to
the introduced ultrametric and discuss the properties of the
corresponding ultrametric spaces.

In Section 5 we construct orthonormal bases of ultrametric
wavelets in the spaces of quadratically integrable functions on
the ultrametric spaces under consideration.

In Section 6 we introduce pseudodifferential operators acting on
complex valued functions on ultrametric spaces, show that these
operators are diagonal in the bases of ultrametric wavelets, and
compute the corresponding eigenvalues.

\section{Directed tree ${\cal T}(X)$ of balls in ultrametric space}

In the present Section we discuss relation between ultrametric
space $X$ and directed tree ${\cal T}(X)$ of balls in $X$. Let us
give the necessary definitions.

\begin{definition}{\sl
An ultrametric space is a metric space with the ultrametric $|xy|$
(where $|xy|$ is called the distance between $x$ and $y$), i.e.
the function of two variables, satisfying the properties of
positivity and non degeneracy
$$
|xy|\ge 0,\qquad |xy|=0\quad \Longrightarrow\quad x=y;
$$
symmetricity
$$
|xy|=|yx|;
$$
and the strong triangle inequality
$$
|xy|\le{\rm max }(|xz|,|yz|),\qquad \forall z.
$$
}
\end{definition}

Consider a complete ultrametric space $X$, satisfying the
following properties:

\medskip

1) The set of all balls of nonzero diameter in $X$ is no more than
countable;

\medskip

2) For any decreasing sequence of balls $\{D^{(k)}\}$,
$D^{(k)}\supset D^{(k+1)}$, diameters of the balls tend to zero;

\medskip

3) Any ball is a finite union of maximal subballs.

\medskip

Property 2 imply the following condition:

For any two balls $I$, $J$ in $X$ (of non zero diameter) and any
sequence of balls $\{D^{(k)}\}$, for which $I\subset
D^{(k)}\subset J$ for all $k$, the sequence $\{D^{(k)}\}$ must be
finite.

\begin{proposition}\label{localcomp}
{\sl Complete ultrametric space, satisfying the properties 1, 2, 3
above, is locally compact. }
\end{proposition}

Here the topology is generated by the ultrametric.

\bigskip

\noindent{\it Proof}\qquad To prove that $X$ is locally compact
(i.e. any ball in $X$ is compact) consider a sequence $\{x_k\}$ in
ball $D$ in $X$. Then, if the ball $D$ is not minimal, it contains
a subball $D'$, which contains infinite subsequence of $\{x_k\}$.
Repeating this procedure, we obtain decreasing sequence of balls
with the diameter tending to zero, where each of the balls
contains infinite subsequence of $\{x_k\}$. Therefore $\{x_k\}$
has a limiting point in $D$. Since the topology on $X$ has a
countable base, this implies the local compactness.

\bigskip

The next proposition gives the example of $\sigma$--additive
measure on $X$.

\begin{proposition}
{\sl Consider on the complete ultrametric space $X$, satisfying
properties 1, 2, 3 the measure $\mu$, satisfying for each ball
$D_I$ the following property:

Measures of all the balls $D_{I_j}$, which are maximal subballs in
$D_I$, are equal.

Then the measure $\mu$ is $\sigma$--additive and has a countable
or finite basis of balls.}
\end{proposition}

\noindent{\it Proof}\qquad It is easy to see that the considered
above condition defines the measure $\mu$ up to multiplication by
a constant.

Then $\sigma$--additivity follows from local compactness of $X$
(in the same way as for the Lebesgue measure) \cite{KF}. The
countable or finite basis for $\mu$ is given by no more than
countable set of balls in $X$. This finishes the proof of the
proposition.

\bigskip

The next definition is the standard definition of a partially
ordered set.

\begin{definition}{\sl Partially ordered set is a set with partially defined order, i.e.
for an arbitrary pair $(x,y)$ of elements of the set these
elements $x$, $y$ either incomparable or $x>y$, or $x<y$, and,
moreover:

1) An element can not be greater or smaller than itself;

2) If $x>y$ and $y>z$, then $x>z$ (transitivity). }
\end{definition}

If any two elements of the partially ordered set are comparable,
then this set is called completely ordered.

A supremum ${\rm sup}\, S$ of the subset $S$ in the partially
ordered set is a minimal element of the partially ordered set,
which is greater or equal to all the elements of the subset $S$.
If any finite subset of the partially ordered set has the unique
supremum, then this partially ordered set is called directed (and
the partial order is called a direction).

\begin{definition}{\sl
A graph is a pair of sets $\{I\}$ (the set of vertices), $\{i\}$
(the set of links), where each of these sets is finite or
countable. Moreover, each link $i$ is a pair of different vertices
$I_0$, $I_1$ (called the beginning and the end of the link). }
\end{definition}

An infinite path in a graph is an injection of the set of natural
numbers into the graph, such that numbers with the difference one
correspond to the neighbor (connected by a link) vertices of the
graph. A path of the length $N$ in a graph is an injection of the
set $1,\dots,N$ into the graph, such that the numbers with the
difference one correspond to the neighbor (connected by a link)
vertices of the graph. In the following we will not specialize,
finite or infinite path we consider if this is clear from the
context. By the applied definition all the considered paths have
an orientation.

The image of 1, defined by the path in the graph, we will call the
beginning of the path. For the finite path of length $N$ the
vertex, which is the image of $N$, will be called the end of the
path. A graph is called connected if for arbitrary two vertices
$I$, $J$ there exists a finite path with the beginning in $I$ and
the end in $J$. A cycle in the graph is a pair of different finite
paths, for which the beginnings and the ends coincide.

A tree is a connected graph without cycles. Consider an arbitrary
tree ${\cal T}$ (finite or infinite), such that the number of
links incident to any of the vertices is finite.

Assume that the tree ${\cal T}$ is partially ordered. If non
maximal vertex $I$ is incident to $p_I+1$ links, we will say that
branching index of the vertex $I$ is equal to $p_I$. If maximal
vertex $I$ is incident to $p_I$ links, we will say that branching
index of the vertex $I$ is equal to $p_I$. Equivalently, branching
index of a vertex $I$ in partially ordered tree is the number of
vertices $I_j$: $I_j<I$, $|II_j|=1$, where the distance between
vertices of the tree is the number of edges in the path connecting
these vertices.

For partially ordered trees ${\cal T}$ consider the following
property:

\bigskip

\noindent{\bf Property 1}\qquad {\sl  In any finite path there
exists the unique maximal vertex.}

\bigskip

In particular, all neighbor (connected by a link) vertices are
comparable.

\bigskip

For ultrametric space $X$ consider the set ${\cal T}(X)$, which
contains all the of balls in $X$ of nonzero diameter, and the
balls of zero diameter which are maximal subbals in balls of
nonzero diameter. On the set ${\cal T}(X)$ there is the natural
partial order: $I<J$ if for the corresponding balls $I\subset J$.
Since for any two balls in an ultrametric space there exists the
unique minimal ball, containing these two balls (which the
supremum of the balls), the mentioned partial order will be a
direction.

The set ${\cal T}(X)$ has the following structure of graph:
vertices of the graph ${\cal T}(X)$ are balls in $X$, two vertices
$I$ and $J$ are connected by a link if one is a subball in the
other, say $I\supset J$, and between $I$ and $J$ there are no
other elements of ${\cal T}(X)$ (i.e. $J$ is a maximal subball in
$I$).

Prove that the graph ${\cal T}(X)$ will be a tree with finite
branching indices. The properties (1), (3) of the ultrametric
space $X$ imply that ${\cal T}(X)$ has no more than countable
number of vertices, and all branching indices are finite.

Since ${\cal T}(X)$ is directed and by the property (2) the graph
${\cal T}(X)$ is connected.

Assume we have a cycle in the graph ${\cal T}(X)$. Take a minimal
element $I$ of the cycle, and consider the balls $J$, $K$ in $X$
lying at the cycle and $I\subset J$, $I\subset K$. By
ultrametricity of the space $X$, for any three balls $I$, $J$, $K$
in $X$, such that $I\subset J$, $I\subset K$, the balls $J$ and
$K$ will be comparable: either $J\subset K$ or $K\subset J$.
Therefore the cycle does not exist and the graph ${\cal T}(X)$ is
a tree.

Then, it is easy to see that the branching index of the tree
${\cal T}(X)$ can not be equal to 1 for any vertex, and for the
maximal vertex (if it exists) the branching index can not be equal
to zero (if $X$ contains at least two points). Moreover, balls of
nonzero diameter in $X$ correspond to vertices of branching index
$\ge 2$ in ${\cal T}(X)$, and the balls of zero diameter which are
maximal subbals in balls of nonzero diameter correspond to
vertices of branching index 0 in ${\cal T}(X)$.

Let us prove that the direction in ${\cal T}(X)$ satisfies the
Property 1. Consider $I$, $J$ in ${\cal T}(X)$ and $K={\rm
sup}(I,J)$. Consider the paths $IJ$, $IK$, $JK$ and the vertex
$$
L=IJ\bigcap IK\bigcap JK
$$
Then $L\ge I$, $L\ge J$ and $L\le K$ which implies $L=K$, i.e. the
supremum $K$ lies at the path $IJ$. Moreover, it is easy to see
that the paths $IK$ and $JK$ are completely ordered, which implies
the Property 1.

We proved the following theorem.

\begin{theorem}\label{the4}
{\sl The set ${\cal T}(X)$ which contains all the of balls of
nonzero diameter, and the balls of zero diameter which are maximal
subbals in balls of nonzero diameter in a non--trivial (containing
at least two points) ultrametric space $X$, satisfying properties
(1), (2), (3) above with the partial order, defined by inclusion
of balls, is a directed tree where all neighbor vertices are
comparable.

Branching index for vertices of this tree may take finite integer
non--negative values not equal to one, and the maximal vertex (if
exists) has the branching index $\ge 2$. Balls of nonzero diameter
in $X$ correspond to vertices of branching index $\ge 2$ in ${\cal
T}(X)$, and the balls of zero diameter which are maximal subbals
in balls of nonzero diameter correspond to vertices of branching
index 0 in ${\cal T}(X)$.

Moreover, the direction in ${\cal T}(X)$ satisfies the Property
1.}
\end{theorem}

Consider the set $X\bigcup {\cal T}(X)$, where we identify the
balls of zero diameter from ${\cal T}(X)$ with the corresponding
points in $X$. We call ${\cal T}(X)$ the tree of balls in $X$, and
$X\bigcup {\cal T}(X)$ the extended tree of balls. One can say
that $X\bigcup {\cal T}(X)$ is the set of all the balls in $X$, of
nonzero and zero diameter.

Introduce the structure of a directed set on $X\bigcup {\cal
T}(X)$. At the tree ${\cal T}(X)$ this structure is already
defined, and the relations of order with points of $X$ are
introduced as follows.

\begin{definition}\label{LobOrd}{\sl Any two points of the ultrametric space $X $ are incomparable.
The relation of order between the points of $X$ and vertices of
the tree ${\cal T}(X)$ are defined as follows: if $x\in X$ and
$I\in {\cal T}(X)$, then $x\le I$ if and only if $x\in I$. }
\end{definition}

This implies the following lemma.

\begin{lemma}{\sl The partial order on $X\bigcup {\cal
T}(X)$, introduced in definition \ref{LobOrd}, is a direction.
This direction can be described as follows.

The supremum
$$
{\rm sup}(x,y)=I
$$
of points $x,y\in X$ is the minimal ball $I$, containing the both
points.

Analogously, for $J\in {\cal T}(X)$ and $x\in X$ the supremum
$$
{\rm sup}(x,J)=I
$$
is the minimal ball $I$, which contains the ball $J$ and the point
$x$. }
\end{lemma}

This construction extends the notion of supremum of two vertices
of the tree ${\cal T}(X)$ (since the ball $I={\rm sup}(J,K)$
corresponds to the minimal ball $I$, which contains the balls $J$
and $K$).

\bigskip

Introduce the structure of an ultrametric space on the extended
tree $X\bigcup {\cal T}(X)$.

\begin{definition}{\sl
For $I$, $J$ in $X\bigcup {\cal T}(X)$ and $I\ne J$ the distance
$|IJ|$ is the diameter of the ball $K\in X\bigcup {\cal T}(X)$,
$K={\rm sup}(I,J)$, for $I$ in $X\bigcup {\cal T}(X)$ the distance
$|II|=0$. }
\end{definition}

In particular, the tree ${\cal T}(X)$ will be (in general,
incomplete) ultrametric space.

Moreover, a measure $\nu$ on ultrametric space $X$ induces the
measure (which we also denote by $\nu$) on the extended tree
$X\bigcup {\cal T}(X)$: the measure of $I\in {\cal T}(X)$ we put
equal to the measure of the corresponding ball in $X$.

\section{Ultrametric on directed trees}

In the present Section we discuss ultrametric on directed trees.
The next theorem describe a family of equivalent ultrametrics on a
directed tree (we call two metrics on the same space equivalent,
if the both define the same set of balls).

\begin{theorem}\label{isultrametric}{\sl If $F(I)$ is a positive
increasing function on a directed tree ${\cal T}$, then the
formula \be\label{eum}|AB|=F({\rm sup}(A,B)),\quad A\ne B,\quad
|AA|=0,\quad \forall A,B\in{\cal T} \ee defines on the tree ${\cal
T}$ the ultrametric (i.e. it is non negative, equal to zero only
for $A=B$, symmetric with respect to permutation of $A$ and $B$,
and satisfies the strong triangle inequality):
$$
|AB|\le {\rm max }(|AC|,|BC|),\qquad \forall A,B,C\in {\cal T}
$$
All the ultrametrics defined in this way are equivalent. }
\end{theorem}

\noindent{\it Proof}\qquad To prove that $|AB|$ is an ultrametric,
it is sufficient to prove that $|AB|$ satisfies the strong
triangle inequality (the other conditions, necessary for
ultrametricity, are obvious).

Consider vertices $A$, $B$, $C$. Let $I={\rm sup}(A,B)$, $J={\rm
sup}(B,C)$, $K={\rm sup}(A,C)$.

Since the both vertices $I$ and $K$ are larger than $A$, these
vertices are comparable (otherwise we would have in the tree
${\cal T}$ the cycle consisting of the two different paths $AI{\rm
sup}(I,K)$, $AK{\rm sup}(I,K)$). Analogously, vertices $I$ and $J$
are comparable; also vertices $J$ and $K$ are comparable.
Therefore $I$, $J$, $K$ is a completely ordered set.

Two variants are possible: either $I=J=K$, or there are some non
coinciding vertices in this set. If $I=J=K$, then by (\ref{eum})
$$
|AB|=|AC|=|BC|
$$
and the strong triangle inequality is satisfied.

Take $I>J$, i.e. ${\rm sup}(A,B)>{\rm sup}(B,C)$. Then ${\rm
sup}(A,C)={\rm sup}(A,B)$, i.e. $I=K$. Thus by (\ref{eum})
$$
|AB|=|AC|>|BC|
$$
and the strong triangle inequality is satisfied.

Analogously, with other choices of order on the set $I$, $J$, $K$
we obtain the strong triangle inequality.

Since the structure of the set of balls, defined in this way,
depends only on the direction on the tree and does not depend on
the function $F$, ultrametrics, defined by different $F$, are
equivalent.

This finishes the proof of the theorem.

\begin{lemma}\label{p1dir}{\sl Partial order in a tree ${\cal T}$,
satisfying Property 1, is a direction.  }
\end{lemma}

\noindent{\it Proof}\qquad Let us prove that, if Property 1 is
satisfied, then for an arbitrary finite set of vertices there
exists the unique supremum (i.e. the partial order is a
direction). It is sufficient to prove this statement for the case
of a pair of vertices, since if for any two vertices $A$, $B$
there exists the unique supremum ${\rm sup}(A,B)$, and we choose
an arbitrary vertex $C$, then
$$
{\rm sup}(A,B,C)={\rm sup}(C,{\rm sup}(A,B))
$$
Therefore the existence of the unique supremum for any pair of
vertices implies the existence and uniqueness of ${\rm
sup}(A,B,C)$. Analogously one can argue for the case of arbitrary
finite number of vertices.

Property 1 implies that for the maximal vertex $C$ at the path
$AB$ the paths $AC$ and $CB$ are completely ordered sets.

For vertices $A$, $B$ we consider the path $AB$ which connects
these vertices and prove, that the unique maximal vertex $C$ at
this path is the supremum for $A$, $B$, and, moreover, this
supremum is uniquely defined.

Let $D$ be some vertex, greater than $A$, $B$. Consider the path
which connects $A$ and $D$. By Property 1 this path contains the
maximal vertex, which greater than $A$ and $D$. Since $D>A$, this
maximal vertex coincides with $D$. Analogously, vertex $D$ is the
maximal vertex at the path which connects $B$ and $D$. Moreover,
the paths $AD$ and $BD$ are completely ordered.

Consider now the paths $AB$, $AD$, $BD$. Since we consider a tree,
there exists the unique vertex $E$, which belongs to all the three
paths. Vertex $E$ satisfies the inequalities
$$
E>A,\qquad E>B,\qquad  E<D.
$$
Therefore $E=C$. This shows that $C$ is the unique supremum for
$A$ and $B$, and the partial order is a direction.

This finishes the proof of the lemma.

\begin{definition}{\sl
We call two paths in the tree ${\cal T}$ equivalent, if they
coincide starting from some vertex (and therefore either both are
infinite or finish at the same vertex). The path from the
equivalence class $x$, which begins in vertex $A$, we denote $Ax$.
}
\end{definition}

The next lemma shows that there exists one--to--one correspondence
between directions in the tree ${\cal T}$, satisfying the Property
1, and equivalence classes of paths in ${\cal T}$. This gives the
constructive way to describe directions on trees, satisfying the
Property 1.

\begin{lemma}{\sl Let us fix some equivalence class $x$ of paths in the tree ${\cal
T}$. Introduce the following partial order in the tree ${\cal T}$.
We say that $A<B$, where $A$, $B$ are vertices of the tree ${\cal
T}$, if there exists the path from the equivalence class $x$, such
that $A$, $B$ lie at this path, and $A<B$ in the sense of the
order at the path (any path is a completely ordered set in the
natural sense: a vertex is smaller if it closer to the beginning
of the path).

This partial order satisfies the Property 1 and therefore is a
direction in ${\cal T}$.

Moreover, any direction in the tree ${\cal T}$, satisfying the
Property 1, can be defined by this procedure (i.e. for any
direction of this kind there exists some equivalence class $x$ of
paths which defines this direction). }
\end{lemma}

\noindent{\it Proof}\qquad Let vertices $A$, $B$ be connected by a
link. Then, since the tree ${\cal T}$ does not contain cycles,
either $B\in Ax$, or $A\in Bx$, and vertices $A$, $B$ are
comparable.

Now let $A$, $B$ be arbitrary vertices. Consider the paths $Ax$,
$Bx$. These paths and the path $AB$ intersect in the unique vertex
$C$. By construction $C>A$, $C>B$, and $C$ is the unique maximal
vertex at the path $AB$. Therefore the introduced partial order
satisfies the Property 1 and by lemma \ref{p1dir} is a direction.

Conversely, consider a direction on the tree ${\cal T}$, which
satisfies the Property 1. Take vertex $A$ and consider the
neighbor vertices (connected to $A$ by links). By the Property 1
this set of vertices (which contain $A$ and the neighbor vertices)
contains the unique maximal vertex $A_1$. Again, consider the set
of vertices, which are the neighbors of $A_1$ (and $A_1$ itself),
and take the maximal vertex $A_2$. Repeating this procedure, we
get the increasing path $AA_1A_2\dots$ (which may be finite or
infinite).

Starting from some vertex $B$, in the analogous way we obtain the
increasing path $BB_1B_2\dots$. Since by the Property 1 the path
$AB$ contains the unique maximal vertex $C={\rm sup}(A,B)$, then
$$
AB\subset AA_1A_2\dots \bigcup BB_1B_2\dots
$$
and the paths $AA_1A_2\dots$, $BB_1B_2\dots$ coincide starting
from vertex $C$. Thus these two paths are in the same equivalence
class.

This finishes the proof of the lemma.

\bigskip

Introduce now the ultrametric on the directed tree ${\cal T}$
(where the direction satisfies the Property 1), which we call the
standard. This example of ultrametric was discussed in \cite{IZV}.

Let us put into correspondence to a link of the tree ${\cal T}$
the branching index of the largest vertex of the link (this
definition is correct since any two vertices, connected by a link,
are comparable). Link is increasing, if the end of the link is
larger than the beginning, and is decreasing in the opposite case.

\begin{definition}\label{ultrametric}{\sl
Consider the directed tree ${\cal T}$, where the direction
satisfies the Property 1. Let all the branching indices of
vertices of the tree are finite and not equal to one, the maximal
vertex either does not exist or has branching index $\ge 2$. Fix
an arbitrary vertex $R$ of the tree (we will call this vertex the
root of the tree).

Let $A$, $B$, $A\ne B$ be vertices of the tree ${\cal T}$ and
$I={\rm sup}(A,B)$, let $R$ be the root of the tree ${\cal T}$.
Let us define the distance $|AB|$ between vertices of the tree as
the product of branching indices of the links along the finite
directed path $RI$
$$
RI=I_0\dots I_N,\qquad I_0=R,\quad I_N=I
$$
in the degrees $\pm 1$, where the branching indices of the
increasing links $I_{j}I_{j+1}$ are in the degree $+1$, and the
branching indices of decreasing links are in the degree $-1$:
\be\label{distance3} |AB|=\prod_{j=0}^{N-1}
p_{I_{j}I_{j+1}}^{\varepsilon_{I_{j}I_{j+1}}}\ee where
$\varepsilon_{I_{j}I_{j+1}}=1$ for $I_{j}<I_{j+1}$,
$\varepsilon_{I_{j}I_{j+1}}=-1$ for $I_{j}>I_{j+1}$.

If vertices $A$, $B$ coincide then the distance between them we
put equal to zero. }
\end{definition}

By theorem \ref{isultrametric} the introduced distance is an
ultrametric.

\section{Absolute and completed tree}

In the present Section we consider completions of directed trees
with respect to ultrametric, defined by theorem
\ref{isultrametric} of the previous Section. This gives a
constructive way to build ultrametric spaces. This construction is
similar to the construction of real (and $p$--adic) numbers by
completion of rational numbers.

\begin{definition}\label{DefX}{\sl For the directed tree ${\cal
T}$ (where the direction satisfies the Property 1) consider the
set $\widetilde {X({\cal T})}$, which is the completion of ${\cal
T}$ with respect to the ultrametric, defined by definition
\ref{ultrametric}. The space $\widetilde {X({\cal T})}$ we call
the completed tree, corresponding to the tree ${\cal T}$.

Consider also the set $X({\cal T})=\widetilde {X({\cal
T})}\backslash \left({\cal T}\backslash {\cal T}_{\rm min}\right)$
(where ${\cal T}_{\rm min}$ is the set of minimal vertices in
${\cal T}$), i.e. $X({\cal T})$ is $\widetilde {X({\cal T})}$,
where all the vertices of the tree ${\cal T}$ are subtracted,
besides the minimal vertices. The space $X({\cal T})$ we call the
absolute of the tree ${\cal T}$. }
\end{definition}

Sets $\widetilde {X({\cal T})}$ and $X({\cal T})$ are complete
ultrametric spaces.

Here we understand subtraction of the tree from its completion as
follows: vertex $A$ is identified with the equivalence class of
sequences of vertices, coinciding with $A$, starting from some
element.

Definition \ref{DefX} is equivalent to the standard definition of
the absolute of the tree, see \cite{Serre} (more definitely, to
the absolute without one point, corresponding to the equivalence
class od increasing paths). Let us call a path in the tree
infinitely continued, if the path is either infinite or finish at
vertex with branching index 0.

\begin{lemma}{\sl The set $\widetilde X({\cal T})$ is in one to one correspondence with the set of classes of
equivalence of decreasing paths in ${\cal T}$. The set $X({\cal
T})$ is in one to one correspondence with the set of classes of
equivalence of infinitely continued decreasing paths in ${\cal
T}$. }
\end{lemma}

\noindent{\it Proof}\qquad We put into correspondence to a
decreasing path in the tree ${\cal T}$ the sequence of vertices as
follows: if the path is infinite, the sequence is the sequence of
vertices at the path; if the path is finite, the sequence is the
sequence of vertices at the path, extended by the infinite
sequence where the term is the last vertex of the path.

Applying this construction to a class of equivalence of decreasing
paths, we get a set of fundamental sequences from the same
equivalence class with respect to the ultrametric (\ref{eum}).
Thus the set of equivalence classes of decreasing paths is a
subset in $\widetilde X({\cal T})$.

Formula (\ref{eum}) implies that any fundamental sequence in
${\cal T}$ with metric (\ref{eum}) contains a subsequence which
coincide, starting from some term, with a subsequence of the
sequence, defined by some decreasing path. Therefore, the set
$\widetilde X({\cal T})$ is equivalent to the set of all
equivalence classes of decreasing paths.

Analogously, an equivalence class of infinitely continued
decreasing paths corresponds to some point of the absolute
$X({\cal T})$, and any equivalence class of sequences of vertices
of the tree, corresponding to some point of the absolute, contains
an equivalence class of infinitely continued decreasing paths.
This finishes the proof of the lemma.

\bigskip

The introduced completed tree $\widetilde X({\cal T})$ coincides
with $X\bigcup {\cal T}(X)$, where $X=X({\cal T})$.  This shows
the duality between complete ultrametric spaces, satisfying
conditions (1), (2), (3) of Section 2, and directed trees with
finite branching index $\ne 1$, the direction satisfying the
Property 1, and where the maximal vertex (if exists) has branching
index $\ge 2$.

\section{Ultrametric wavelets}

In the present and next Sections we consider complete ultrametric
space $X$, satisfying the properties 1, 2, 3. Consider a
$\sigma$--additive possessing a countable or finite basis of balls
positive measure $\nu$ on ultrametric space $X$.

Build a basis in the space $L^2(X,\nu)$ of quadratically
integrable with respect to the measure $\nu$ functions, which we
will call the basis of ultrametric wavelets.

Denote $V_{I}$ the space of functions on $X$, generated by
characteristic functions of the maximal subballs in the ball of
nonzero radius $D_I$. Correspondingly, we denote $V^0_{I}$ the
subspace of codimension 1 in $V_{I}$ of functions with zero mean
with respect to the measure $\nu$. The proof of the following
lemma is straightforward.

\begin{lemma}\label{orthogonality}{\sl
Spaces $V^0_{I}$ for different $I$ are orthogonal in $L^2(X,\nu)$.
}
\end{lemma}

We introduce in the space $V^0_{I}$ some orthonormal basis
$\{\psi_{Ij}\}$, where the number of vectors in the basis is
obviously less or equal to $p_I-1$. The next theorem shows how to
construct the orthonormal basis in $L^2(X,\nu)$, taking the union
of bases $\{\psi_{Ij}\}$ in spaces $V^0_{I}$ over all non minimal
$I$.

\begin{theorem}\label{basisX}
{\sl 1) Let the ultrametric space $X$ contains an increasing
sequence of embedded balls with infinitely increasing measure.
Then the set of functions $\{\psi_{Ij}\}$, where $I$ runs over all
non minimal vertices of the tree ${\cal T}(X)$ is an orthonormal
basis in $L^2(X,\nu)$.

2) Let for the ultrametric space $X$ there exists the supremum of
measures of the balls, which is equal to $A$. Then the set of
functions $\{\psi_{Ij}, A^{-{1\over 2}}\}$, where $I$ runs over
all non minimal vertices of the tree ${\cal T}(X)$ is an
orthonormal basis in $L^2(X,\nu)$.
 }
\end{theorem}

The introduced in the present theorem basis we call the basis of
ultrametric wavelets. For the case when the measure $\nu$ is
defined in the special way: the measure of a ball is equal to its
diameter, this theorem (and the results of the next Section on
diagonalization of ultrametric PDO) was obtained in \cite{IZV}.

\bigskip

\noindent{\it Proof}\qquad By lemma \ref{orthogonality} the
described in the statement of the theorem functions are
orthonormal.

To prove the totality we use the Parseval identity. Since the set
of characteristic functions of all the balls $D_I$ is total in
$L^2(X,\nu)$, to prove the totality it is sufficient to prove the
Parseval identity only for characteristic functions $\chi_I(x)$.

Consider the characteristic function $\chi_I$ of the ball $D_I$,
satisfying the condition $\nu(D_I)>0$ (and therefore $\chi_I\ne 0
$ in $L^2(X,\nu)$). Expand the characteristic function $\chi_I(x)$
over the wavelets. It is sufficient to consider wavelets
$\psi_{Jj}$ with $J>I$.

Denote $P_{V}$ the orthogonal projection onto $V$ in $L^2(X,\nu)$.
Then
$$
P_{V^0_{J}}=P_{V_{J}}-P_{\chi_{J}}
$$
Consider the vector
$$
\widetilde \chi_I=\sum_{J\in{\cal T}}
P_{V^0_{J}}\chi_I=\sum_{J>I}\left(P_{V_{J}}-P_{\chi_{J}}\right)\chi_I=
\sum_{J>I}\left(P_{\chi_{J-1,I}}-P_{\chi_{J}}\right)\chi_I
$$
where $(J-1,I)$ is the (uniquely defined) maximal vertex, which is
less than $J$ and larger than $I$. The above vector is expanded
into the series over the orthogonal vectors. Compute the square of
the length $\widetilde \chi_I$:
$$
\|\widetilde
\chi_I\|^2=\sum_{J>I}\|\left(P_{\chi_{J-1,I}}-P_{\chi_{J}}\right)\chi_I\|^2
$$
Since for $J\ge I$
$$
P_{\chi_{J}}\chi_I={\nu(D_I)\over\nu(D_J)}\chi_J
$$
(this expression is correct since $\nu(D_J)>0$ for $J\ge I$), we
get
$$
\|\left(P_{\chi_{J-1,I}}-P_{\chi_{J}}\right)\chi_I\|^2=
\|{\nu(D_I)\over\nu(D_{J-1,I})}\chi_{J-1,I}-{\nu(D_I)\over\nu(D_J)}\chi_J\|^2=
$$
$$
=\left({\nu(D_I)\over\nu(D_{J-1,I})}-{\nu(D_I)\over\nu(D_J)}\right)^2\nu(D_{J-1,I})+
\left({\nu(D_I)\over\nu(D_{J})}\right)^2(\nu(D_{J})-\nu(D_{J-1,I}))=
$$
$$
=\nu^2(D_I)\left[{1\over\nu(D_{J-1,I})}-{1\over\nu(D_{J})}\right]
$$
This implies \be\label{pars1} \|\widetilde
\chi_I\|^2=\nu^2(D_I)\sum_{J>I}\left[{1\over\nu(D_{J-1,I})}-{1\over\nu(D_{J})}\right]
=\nu^2(D_I)\lim_{J\to\infty,J>I}
\left[{1\over\nu(D_{I})}-{1\over\nu(D_J)}\right] \ee where the
limit at the RHS is the limit of the expression in square brackets
for the sequence of increasing $J$, which begins from $I$.

In the case 1 formula (\ref{pars1}) implies the Parseval identity
for $\chi_I$.

In the case 2 we get for (\ref{pars1})
$$
\nu^2(D_I) \left[{1\over\nu(D_{I})}-{1\over A}\right]
$$
Since in this case to prove the totality we have to add to the
expression above the term corresponding to the contribution of the
normed constant:
$$
{\nu^2(D_J)\over A}
$$
we again obtain the Parseval identity, which finishes the proof of
the theorem.

\section{Diagonalization of ultrametric PDO}

In the present Section we study the ultrametric pseudodifferential
operator (or the PDO) of the form
$$
Tf(x)=\int T^{({\rm sup}(x,y))}(f(x)-f(y))d\nu(y)
$$
Here $T^{(I)}$ is some complex valued function on the tree ${\cal
T}( X)$. Thus the structure of this operator is determined by the
direction on $X\bigcup {\cal T}(X)$.

\begin{theorem}\label{04}{\sl Let the following series converge absolutely:
\be\label{seriesconverge} \sum_{J>R} T^{(J)}
(\nu(D_J)-\nu(D_{J-1,R})) <\infty \ee Then the operator
$$
Tf(x)=\int T^{({\rm sup}(x,y))}(f(x)-f(y))d\nu(y)
$$
has the dense domain in $L^2(X,\nu)$, and is diagonal in the basis
of ultrametric wavelets from the theorem \ref{basisX}:
\be\label{lemma2.1} T\psi_{Ij}(x)=\lambda_I \psi_{Ij}(x) \ee with
the eigenvalues: \be\label{lemma4} \lambda_{I}=T^{(I)}
\nu(D_I)+\sum_{J>I} T^{(J)} (\nu(D_J)-\nu(D_{J-1,I})) \ee and is
self--adjoint if $T^{(I)}$ is real valued function.

Here $(J-1,I)$ is the maximal vertex which is less than $J$ and
larger than $I$.

Also the operator $T$ kills constants. }
\end{theorem}

\noindent{\it Proof}\qquad Consider the action of the operator
onto the wavelet $\psi_{Ij}$:
$$
T\psi_{Ij}(x)=\int T^{({\rm
sup}(x,y))}\left(\psi_{Ij}(x)-\psi_{Ij}(y) \right)d\nu(y)
$$

Consider the following cases.

1) Let $x$ does not lie at the ball $D_I$. Then
$$
T\psi_{Ij}(x)=-T^{({\rm sup}(x,I))}\int \psi_{Ij}(y)d\nu(y)=0
$$

2) Let $x\in D_I$. Denote $\mu(D_I)$ the diameter of the ball
$D_I$. Then
$$
T\psi_{Ij}(x)= \left(\int_{|xy|>\mu(D_I)}+ \int_{|xy|=\mu(D_I)}+
\int_{|xy|<\mu(D_I)}\right)T^{({\rm sup}(x,y))}
(\psi_{Ij}(x)-\psi_{Ij}(y))d\nu(y)=
$$
$$
= \left(\int_{|xy|>\mu(D_I)}+ \int_{|xy|=\mu(D_I)}\right)T^{({\rm
sup}(x,y))} (\psi_{Ij}(x)-\psi_{Ij}(y))d\nu(y)=
$$
$$
= \psi_{Ij}(x)\int_{|xy|>\mu(D_I)} T^{({\rm sup}(x,y))}d\nu(y)+
\int_{|xy|=\mu(D_I)}T^{({\rm sup}(x,y))}
(\psi_{Ij}(x)-\psi_{Ij}(y))d\nu(y)=
$$
$$
= \psi_{Ij}(x)\int_{|Iy|>\mu(D_I)}T^{({\rm sup}(I,y))}d\nu(y)+
T^{(I)}\nu(D_{I})\psi_{Ij}(x)
$$

To prove the last identity let us compute for $\psi\in V^0_{I}$
the integral
$$
\int_{|xy|=\mu(D_I)}T^{({\rm sup}(x,y))} (\psi(x)-\psi(y))d\nu(y)=
T^{(I)}\int_{|xy|=\mu(D_I)}(\psi(x)-\psi(y))d\nu(y)=
$$
$$
=T^{(I)}\sum_{j=0}^{p_I-1}\chi_{I_j}(x)\left[\int_{D_I\backslash
D_{I_j}}(\psi(x)-\psi(y))d\nu(y)\right]=
$$
$$
=T^{(I)}\sum_{j=0}^{p_I-1}\chi_{I_j}(x)\left[\psi(x)(\nu(D_I)-\nu(D_{I_j}))-\int_{D_I\backslash
D_{I_j}}\psi(y)d\nu(y)\right]=
$$
$$
=T^{(I)}\sum_{j=0}^{p_I-1}\chi_{I_j}(x)\left[\psi(x)(\nu(D_I)-\nu(D_{I_j}))+\int_{
D_{I_j}}\psi(y)d\nu(y)\right]=T^{(I)}\nu(D_I)\psi(x)
$$
Here $D_{I_j}$ are the maximal subballs in $D_{I}$.

We get
$$
T\psi_{Ij}(x)= \lambda_I\psi_{Ij}(x)
$$
where
$$
\lambda_I=T^{(I)}\nu(D_I)+\int_{|Iy|>\mu(D_I)}T^{({\rm
sup}(I,y))}d\nu(y)
$$
For $J>I$
$$
\int_{|Iy|=\mu(D_J)}d\nu(y)=\nu(D_J)-\nu(D_{J-1,I})
$$
Since any two increasing paths in a directed tree coincide
starting from some vertex, condition (\ref{seriesconverge})
provides convergence of the integral $\int_{|Iy|>\mu(D_I)}T^{({\rm
sup}(I,y))}d\nu(y)$.

This implies
$$ \lambda_{I}=T^{(I)} \nu(D_I)+\sum_{J>I} T^{(J)}
(\nu(D_J)-\nu(D_{J-1,I}))
$$

Proof that the operator $T$ kills constants is straightforward.
This finishes the proof of the theorem.

\bigskip

\bigskip

\noindent{\bf Acknowledgements}

\noindent The author is grateful to V.S.Vladimirov, I.V.Volovich,
A.Yu.Khrennikov, V.A.Avetisov and A.Kh.Bikulov for discussions and
important remarks. The author is partially supported by CRDF
(project UM1--2421--KV--02), RFFI (project 02--01--01084), and the
grant of the President of Russian Federation for support of
scientific school N.Sh.1542.2003.1.


\begin{thebibliography}{99}

\bibitem{IZV} A.Yu.Khrennikov, S.V.Kozyrev,
Pseudodifferential operators on ultrametric spaces and ultrametric
wavelets, http://arxiv.org/abs/math-ph/0412062

\bibitem{Vl1} {\it Vladimirov V.S.}, On the spectrum of some pseudodiferential operators
over the field of $p$--adic numbers
// Algebra and Analysis. 1990. v.2 N 6. p.107--124.
\bibitem{VVZ} {\it  Vladimirov V.S.,   Volovich I.V.,  Zelenov Ye.I.}
$p$--Adic Analysis and Mathematical Physics. Singapore: World
Scientific, 1994.
\bibitem{Andr3} {\it A.Yu.Khrennikov} Non--Archimedean Analysis and it Applications.
Moscow: Nauka, Fizmatlit, 2003. (in Russian)
\bibitem{Andr1}  {\it Khrennikov A.} Non--Archimedean Analysis:
Quantum Paradoxes, Dynamical Systems and Biological Models.
Dordrecht: Kluwer Academic Publishers, 1997.
\bibitem{AlKa}   {\it Albeverio S.,   Karwowosky W.} A random walk on
$p$--adic numbers, in ''Stochastic Process--Physics and Geometry
II '' (S. Albeverio, U. Cattaneo, D. Merlini, Eds.), Proc. Locarno
(1991), pp.61--74. Singapore: World Scientific, 1995.
\bibitem{Kochubei1}   {\it Kochubei A.N.} Pseudo--Differential Equations and Stochastics over
Non--Archimedean Fields. New York: Marcel Dekker, 2001.
\bibitem{Kochubei2}
{\it  Kochubei A.N.} Fundamental solutions of pseudodifferential
eqiations, related to $p$--adic quadratic forms // Izvestia
Academii Nauk Seria Math. 1998. V. 62. N 6. P. 103--124.

\bibitem{wavelets}  {\it Kozyrev S.V.} Wavelet analysis as a
$p$--adic spectral analysis // Russian Math. Izv. 2002. V. 66. N
2. P. 367. http://arxiv.org/abs/math-ph/0012019

\bibitem{Benedetto} J.J.Benedetto, R.L.Benedetto, A wavelet theory
for local fields and related groups, The Journal of Geometric
Analysis, 14 (3), 2004, pp. 423--456

\bibitem{Benedetto1} R.L.Benedetto, Examples of wavelets for local
fields, http://arxiv.org/math.CA/abs/0312038


\bibitem{nhoper} {\it Kozyrev
S.V.} $p$--Adic pseudodifferential operators and $p$--adic
wavelets // Theor. Math. Physics.  2004. V. 138. N 3, P. 322--332.
http://arxiv.org/abs/math-ph/0303045
\bibitem{trudy} S.V. Kozyrev, $p$--Adic pseudodifferential
operators: methods and applications, Trudy MIAN, v. 245, 2004,
p.154--165
\bibitem{0403440} S.V. Kozyrev, V.Al. Osipov, V.A. Avetisov, Non--Degenerate
Ultrametric Diffusion, http://arxiv.org/abs/cond-mat/0403440



\bibitem{ABK}  {\it Avetisov V.A.,  Bikulov  A.H.,  Kozyrev S.V.}
Application of $p$--adic analysis to models of spontaneous
breaking of replica symmetry // J. Phys. A: Math. Gen. 1999. V.
32. P. 8785--8791. http://arxiv.org/abs/cond-mat/9904360
\bibitem{PaSu}  {\it Parisi G.,   Sourlas N.}
$p$--Adic numbers and replica symmetry breaking // European Phys.
J. B. 2000. V. 14. P. 535--542.
http://arxiv.org/abs/cond-mat/9906095

\bibitem{ABKO} {\it Avetisov V.A.,  Bikulov A.H.,  Kozyrev S.V.,  Osipov V.A.}
$p$--Adic Models of Ultrametric Diffusion Constrained by
Hierarchical Energy Landscapes // J. Phys. A: Math. Gen. 2002. V.
35. P. 177--189. http://arxiv.org/abs/cond-mat/0106506

\bibitem{Trudy2}  {\it Avetisov V.A.,  Bikulov A.H.,  Osipov V.A.},
$p$--Adic models of ultrametric diffusion in conformational
dynamics of macromolecules. Trudy MIAN, v. 245, 2004. p.55--65

\bibitem{Vstring} I.V.Volovich, $p$--Adic String,
{\it Class. Quantum Gravity}, 1987, {\bf 4}, P. L83--L87
\bibitem{Freund} L.Brekke, P.G.O.Freund,  M.Olson,  E.Witten,
Non--archimedian string dynamics. Nucl. Physics, B302 (1988)
P.365-–402.


\bibitem{Andr2}  {\it Khrennikov A.} Classical and quantum mental models and Freud's theory of unconscious
mind. V\"axj\"o University, V\"axj\"o, Sweden: V\"axj\"o
University press, 2002.

\bibitem{Serre} {\it Serre J.P.} Trees. New York, Berlin: Springer Verlag,
1980.

\bibitem{KF} A.N.Kolmogorov, S.V.Fomin, Elements of theory of
functions and functional analysis, Moscow, Nauka, 1976 (in
Russian)

\end{thebibliography}
\end{document}